\documentclass[preprint,12pt,authoryear]{elsarticle}

\usepackage{amssymb}
\usepackage{latexsym}
\usepackage{amsmath}
\usepackage{color}
\usepackage{graphicx,dblfloatfix}
\usepackage{subfigure}
\usepackage[round]{natbib}
\usepackage[margin=1in]{geometry}
\usepackage{booktabs}
\usepackage{lscape}
\usepackage{lineno}
\usepackage{tabularx}
\usepackage[table]{xcolor}    

\newcommand{\bfm}[1]{\mbox{\boldmath{$#1$}}}

\journal{Icarus}

\begin{document}

\begin{frontmatter}



\title{Rotationally induced failure of irregularly shaped asteroids}


\author{Masatoshi Hirabayashi$^1$ and Daniel J. Scheeres$^2$}

\address{$^1$Department of Aerospace Engineering, Auburn University, 211 Davis Hall, Auburn, AL 36849-5338, United States \\
$^2$Aerospace Engineering Sciences, University of Colorado, 429 UCB, Boulder, CO 80309-0429, United States}

\begin{abstract}
Many asteroids are rubble piles with irregular shapes. While the irregular shapes of large asteroids may be attributed to collisional events, those of small asteroids may result from not only impact events but also rotationally induced failure, a long-term consequence of small torques caused by, for example, solar radiation pressure. A better understanding of shape deformation induced by such small torques will allow us to give constraints on the evolution process of an asteroid and its structure. However, no quantitative study has been reported to provide the relationship between an asteroid's shape and its failure mode due to its fast rotation. Here, we use a finite element model (FEM) technique to analyze the failure modes and conditions of 24 asteroids with diameters less than 30 - 40 km, which were observed at high resolution by ground radar or asteroid exploration missions. Assuming that the material distribution is uniform, we investigate how these asteroids fail structurally at different spin rates. Our FEM simulations describe the detailed deformation mode of each irregularly shaped asteroid at fast spin. The failed regions depend on the original shape. Spheroidal objects structurally fail from the interior, while elongated objects experience structural failure on planes perpendicular to the minimum moment of inertia axes in the middle of their structure. Contact binary objects have structural failure across their most sensitive cross sections. We further investigate if our FEM analysis is consistent with earlier works that theoretically explored a uniformly rotating triaxial ellipsoid. The results show that global shape variations may significantly change the failure condition of an asteroid. Our work suggests that it is critical to take into account the actual shapes of asteroids to explore their failure modes in detail.
\end{abstract}


\end{frontmatter}



\section{Introduction}
\label{Sec:Intro}
Over the last few decades, spacecraft explorations and ground observations have revealed that many small asteroids are loosely packed aggregates, so-called rubble piles, and have irregular shapes. These asteroids are subject to many different kinds of external forces that could change their spin states. Some forces may be small but act continuously, generating significant effects on their rotational and orbital evolution over the lifetime. Such forces include solar radiation pressure. 

Solar radiation pressure generates small but continuous forces on sunlit surfaces of planetary objects. If asteroids are small enough to be affected by solar radiation pressure-driven forces, their asymmetric bodies experience torques and change their spin states \citep{Rubincam2000}. Ground observations have detected rotational acceleration/deceleration of small asteroids due to solar radiation pressure \citep[e.g.][]{Lowry2007, Taylor2007, Durech2008}. The breakup of the active asteroid P2013/R3 has been interpreted as the result of fast rotation caused by solar radiation pressure \citep{Jewitt2014R3, Jewitt2017}. Spin-state variations due to solar radiation pressure, called the YORP effect, depend on an asteroid's orientation of an asteroid towards the Sun \citep{Nesvorny2007} and its shape \citep{Scheeres2007B}, causing complex rotational dynamics \citep{Scheeres2008Rotational}. Topographic sensitivity of the YORP effect plays a significant role in rotational dynamics of small asteroids significantly \citep{Statler2009}. The YORP effect is also responsible for the formation of binary, triples, and pairs \cite[e.g.][]{Cuk2005, Goldreich2009, Jacobson2011}. 

The YORP effect is considered to have caused small asteroids to reach their spin limits. The behavior of these asteroids at the spin limits is key to answering their evolution. Friction affects shape equilibrium \citep{Holsapple2001} while cohesion may help asteroids keep asteroids' original shapes at fast rotation \citep{Holsapple2007}. Surface deformation processes also contribute to material shedding \citep{Scheeres2015Land}. A hard-sphere discrete element model showed that the equatorial ridge of a top-shaped object might result from the movement of materials toward the equator due to fast rotation \citep{Walsh2008, Walsh2012}. Shape deformation changes the YORP-driven torque, causing stochastic variations in the rotational state of an asteroid \citep{Cotto2015}. Soft-sphere discrete element methods have shown that a randomly packed sphere might have internal deformation at fast spin \citep[e.g.][]{Sanchez2012, Sanchez2016} although a heterogeneity in the internal structure would control the failure modes and conditions \citep{Zhang2017}.

Substantial progress has been made in theoretical modeling of the internal deformation processes of asteroids. A key trend is that models assumed an asteroid to be a triaxial ellipsoid. This assumption allows for deriving the internal stress analytically, making problems clear and reasonably solvable \citep[e.g.][]{Love1927, Dobrovolskis1982, Holsapple2001}. While we have seen many pioneering works that shed light on the deformation mechanism of asteroids \citep[e.g.][]{Dobrovolskis1982, Davidsson2001, Holsapple2001, Holsapple2004, Holsapple2007, Holsapple2010, Sharma2009}, we assert that the shape evolution due to rotationally induced deformation is still an open question. The main reason is that asteroids do not have ideal shapes; in other words, they are neither spheres nor ellipsoids. 

The purpose of this study is to use a finite element model (FEM) analysis \citep{Hirabayashi2014DA, Hirabayashi2016Itokawa, Scheeres2016Bennu, Hirabayashi2016Nature} to quantify the failure modes and conditions of 24 observed asteroids of which high-resolution polyhedron shape models were generated. We choose asteroids smaller than 30 - 40 km in diameter because asteroids in this size range could be spun up/down by the YORP effect \citep{Vokrouhlicky2015}. The plastic FEM technique developed by the authors is based on the work done by \cite{Holsapple2008A}. Here, we will investigate the stress conditions of these asteroids at different spin rates and evaluate their failure modes.

We outline the contents employed in this paper. First, we will discuss the strength model used. Second, we will categorize 24 asteroid into four shape types: contact binary objects, elongated objects, spheroidal objects, and non-classified objects. Although this classification is subjective, it will help us quantify the failure modes and conditions of these asteroids. Third, we will review our plastic FEM technique. Fourth, using the FEM technique, we will compute the failure mode and condition of each asteroid. Finally, we will compare the results from our FEM technique and those from earlier works that used a volume-averaging technique to explore a uniformly rotating triaxial ellipsoid \citep[e.g.][]{Holsapple2004, Holsapple2007, Holsapple2010, Sharma2009, Rozitis2014, Hirabayashi2014Biaxial}. We finally note that we distinguish ``failure mode" and ``failure condition." The failure mode means what regions in an asteroid would structurally fail while the failure condition describes when the asteroid experiences such a failure mode. 

\section{Shear resistance of a material}
Our model assumes granular materials in asteroids to be continuum media; in other words, we consider very fine-grained materials filling interstices in the continuum limit, the idea of which is consistent with \cite{Sanchez2011}. This section discusses the strength model used in this study. We use the Drucker-Prager model to describe the yield condition \citep{Chen1988}:
\begin{eqnarray}
f = \alpha I_1 + \sqrt{J_2} - s \le 0, \label{Eq:DPcriterion}
\end{eqnarray}
where $I_1$ and $J_2$ are the stress invariants. $\alpha$ and $s$ are defined such that the Drucker-Prager yield surface touches the Mohr-Coulomb yield surface at the compression meridian in the principal stress space. At the compression meridian, these parameters are given as \citep{Chen1988}
\begin{eqnarray}
\alpha = \frac{2 \sin \phi}{\sqrt{3} (3 - \sin \phi)}, \:\:\: s = \frac{6 Y \cos \phi}{\sqrt{3} (3 - \sin \phi)}, \label{Eq:alpha&s}
\end{eqnarray} 
where $Y$ is the cohesive strength, and $\phi$ is the friction angle. The friction angle of a geological material is widely known to range from 30$^\circ$ to 40$^\circ$ \citep{Lambe1969}. Thus, in this study, we fix the friction angle at the mean value of this range, i.e., 35$^\circ$, which is consistent with earlier works \citep[e.g.][]{Hirabayashi2016Nature}. 

Studies have shown that rubble pile asteroids might have cohesive strength. Wan der Waals forces would generate a significant level of cohesive strength, compared to the gravity level in an asteroid less than kilometers in diameter \citep{Scheeres2010}. The breakup event of P/2013 R3 gave constraints on the cohesive strength of this asteroid as 50 - 250 Pa \citep{Hirabayashi2014R3}. The fast rotation of 1950 DA indicated that this asteroid needs a cohesive strength higher than $\sim$80 Pa, which is consistent with that of P2013 R3 \citep{Rozitis2014, Hirabayashi2014DA}. Recent observations proposed that fast-rotating asteroids would have cohesive strengths higher than 100 Pa \citep{Polishook2016}. Since these studies show that the cohesive strength is a key physical property that characterizes the internal structure, we consider the cohesive strength to be a free parameter to quantify the failure modes and conditions.

\section{Subjective shape classification of asteroids}
We investigate the rotationally induced failure modes and conditions of 24 asteroids with mean diameters less than 30 - 40 km, which were observed at high resolution. The shape models of these asteroids were developed either by high-resolution images from asteroid exploration missions or by ground radar observations. We show the physical properties of these asteroids in Table \ref{Table:AsteroidList}. The first column gives the asteroid name. The second column describes the current spin period of each asteroid. The third column shows the bulk density, and we only describe the well-estimated values; otherwise, we put dashes and assume the bulk density to be 2.5 g/cm$^3$. The fourth column indicates the volume of the shape model. The fifth column shows the shape class. 

Before explaining the details of the shape classification, we introduce the literature of {\it subjective} shape classification, which was mainly based on observations. Asteroids observed at high resolution have usually been categorized into one of the following five classes: contact binaries, elongated bodies, spheroidal bodies, multiple asteroid systems, and non-classified asteroids\footnote{\cite{Taylor2012} called these asteroids irregularly shaped asteroids.} \citep{Taylor2012}. For contact binaries, \cite{Benner2015} provided a clear (but still subjective) description that they consist of two lobes, which might once have been separated, currently resting on each other. More broadly, it can be considered that these objects have narrow necks. Elongated bodies have stretched shapes along one direction without narrow necks. Spheroidal bodies are more or less rounded and thus have relatively low aspect ratios. Multiple asteroid systems are bodies having satellites. Some binary systems consist of a spheroidal primary having an equatorial ridge, known as the ``top" shape, and a relatively small secondary \citep{Margot2015, Walsh2015}. Fourteen percent of NEAs imaged by radar are contact-binary candidates \citep{Taylor2012}, while 16$\%$ of them are binary asteroids \citep{Margot2002}. Lastly, non-classified asteroids are those not categorized into these classes. 

This study denotes the asteroid shapes using four alphabetic letters, C, E, S, and N, omitting multiple system objects. Asteroids are categorized into these four shapes in a subjective way based on the above arguments; therefore, the shapes of some asteroids might not be determined uniquely. Shape C means a contact binary object. There are six contact binaries in our asteroids (see Table \ref{Table:AsteroidList}). For Mithra, based on an uncertainty of its pole direction, \cite{Brozovic2010} reported a prograde model and a retrograde model. Here, we use the prograde model. Shape E indicates an elongated body, including four objects. Shape S is a spheroidal body. More than half of the considered asteroids are spheroidal. The top-shaped asteroids are included in this class. For the shape of 1950 DA, we use the retrograde shape model by following \cite{Farnocchia2014}, who showed a $99 \: \%$ likelihood that this object is a retrograde rotator based on the analysis of non-gravitational perturbation. We also mention that although the shape model of 2000 ET70 was updated \citep{Marshall2016}, we refer to the model developed by \cite{Naidu2013}. Among these objects, 1999 KW4, 1994 CC, 2001 SN263, and 2000 DP107 are multiple system asteroids, and we analyze the current shapes of their primaries. Shape N stands for a non-classified object. Golevka, a tooth-like shape, is the only shape categorized into this type. Note that our alphabetic classification is not related to taxonomy classification \citep{Bus2002, DeMeo2009}. 

\begin{table}
\begin{center}
\caption{Physical properties of 24 asteroids considered in this study. $P$ is the rotation period [hr], $\rho$ is the bulk density [g/cm$^3$], and $V$ is the total volume [m$^3$]. The fifth column shows the shape classification. C stands for contact binary objects, E indicates elongated objects, S means spheroidal objects, and N shows non-categorized objects. We omit errors for spin periods and the volume although omitted errors for spin periods are typically small whereas those for the volume can be large.}
\resizebox{17cm}{!} {
\rowcolors{2}{gray!25}{white}
\begin{tabular}{l  l  l  l  l  l} 
 \rowcolor{gray!50}
\toprule
Asteroid Name & $P$ [hr] & $\rho$ [g$/$cm$^3$] & $V$ [km$^3$] & Shape & Reference \\ 
\hline 
\hline
     (433) Eros                  & 5.270  & $2.67\pm0.03$ & 2533 & C & \cite{MIller2002} \\ & & & & & \cite{Gaskell2008Eros} \\
     (1580) Betulia            & 6.138  & $-$             & 81.98 & S & \cite{Magri2007} \\ & & & & &  \cite{Kaasalainen2004} \\
     (1620) Geographos   & 5.223  & $-$             & 8.868 & E & \cite{Magnusson1996} \\ & & & & &  \cite{Ostro1996} \\ & & & & &  \cite{RHudson1999} \\ 
     (2063) Bacchus         & 14.90   & $-$            & 0.1313 & E & \cite{Benner1999}  \\ 
     (2100) Ra-Shalom     & 19.79  & $-$             & 6.203 & S & \cite{MShepard2008}  \\ 
     (4179) Toutatis          & 175.2   & $-$             & 7.670 & C & \cite{RHudson1995} \\ & & & & &  \cite{Spencer1995} \\ 
     (4660) Nereus           & 15.16 & $-$ & 0.01940 & E &\cite{Brozovic2009}  \\ 
     (4486) Mithra            & 67.5 & $-$ & 2.532 & C & \cite{Brozovic2010} \\
     (4769) Castalia         & 4.07 & $-$ & 0.6678 & C & \cite{Ostro1990} \\ & & & & & \cite{RHudson1994} \\ & & & & & \cite{Scheeres1996} \\
     (6489) Golevka                  & 6.029 & $-$ & 0.07795 & N & \cite{RHudson2000}  \\ 
     (8567) 1996 HW1              & 8.762  & $-$ & 4.336 & C & \cite{Magri2011}  \\ 
     (10115) 1992 SK                & 7.318  & $-$ & 0.5314 & E & \cite{Busch2006}  \\ 
     (25143) Itokawa                 & 12.13   & $1.9\pm0.13$ & 0.01760 & C & \cite{Fujiwara2006} \\      
     (29075) 1950 DA                & 2.122  & $1.7\pm0.7$  & 1.145 & S & \cite{Busch2007} \\ & & & & & \cite{Rozitis2014} \\
     (33342) 1998 WT24           & 3.697 & $-$ & 0.03751 & S & \cite{Busch2008}  \\ 
     (52760) 1998 ML14           & 14.83  & $-$ & 0.5112 & S & \cite{Ostro2001}  \\ 
     (66391) 1999 KW4   & 2.765 &  $1.97\pm0.24$ & 1.195 & S & \cite{Ostro2006} \\ 
     (101955) Bennu                 & 4.297 & $1.26\pm0.07$ & 0.06227 & S & \cite{Nolan2013} \\ & & & & & \cite{Chesley2014} \\
     (136617) 1994 CC   & 2.389 & $2.6\pm0.6$ & 0.1249 & S & \cite{Fang2011} \\ & & & & & \cite{Brozovic2011}  \\  
     (153591) 2001 SN263       & 3.4256 & $1.1\pm0.2$ & 8.963 & S & \cite{Becker2015} \\
     (162421) 2000 ET70         & 8.96 & $-$ & 6.065 & S & \cite{Naidu2013} \\
     (185851) 2000 DP107       & 2.775 & $1.381 \pm 0.244$ & 0.3369 & S & \cite{Naidu2015} \\
     2002 CE26                        & 3.293  & $0.9 + 0.5/-0.4$ & 21.67 & S & \cite{MShepard2006}  \\ 
     2008 EV5                          &  3.725  & $-$ & 0.03484 & S & \cite{Busch2011}  \\ 
\bottomrule
\end{tabular}
}
\label{Table:AsteroidList}	
\end{center}
\end{table}

\section{Analysis}
\subsection{Normalization and notational definitions}
\label{Sec:Normalization}
We use normalized parameters in this paper. We apply the definitions used by \cite{Hirabayashi2015internal} and \cite{Hirabayashi2015Sphere}. The density, mean radius, and gravitational constant are defined as $\rho$, $R$ and $G$, respectively. We normalize the lengths, forces, spin rates, and stresses (and thus cohesive strength) by $R$, $\pi \rho G R$, $\sqrt{\pi \rho G}$, and $\pi \rho^2 G R^2$, respectively. In our discussion, the normalized spin rate is defined as $\omega$. We also denote the normalized cohesive strength as $Y$. We introduce the key physical parameters used in our work in Table \ref{Table:param}. 

We introduce four physical parameters to discuss the failure modes of the sample asteroids. The first parameter is the minimum cohesive strength, which is denoted as $Y^\ast$. This quantity represents the minimum value of the cohesive strength that can hold the original shape of the body. The second parameter is the normalized current spin rate, which is given as $\omega_c$. The third parameter, $\omega_0$, describes the normalized critical spin rate at which stresses transit from a compression-dominant mode to a tension-dominant mode. Thus, tension starts to control the failure mode at a spin rate higher than $\omega_0$. The fourth parameter, $\eta$, shows a factor of the increase in $Y^\ast$ at a high spin rate. $\omega_0$ and $\eta$ will be discussed more in Section \ref{Sec:FMD}.  

\begin{table}
\begin{center}
\caption{Notational definitions of key parameters used in this work.}
\label{Table:param}
\rowcolors{2}{gray!25}{white}
\begin{tabularx}{\textwidth}{X l l}
 \rowcolor{gray!50}
\toprule
Parameters & Symbols & Units \\
\hline
\hline
Gravitational constant $(=6.674 \times 10^{-11})$ & $G$ & m$^3$ kg$^{-1}$ s$^{-2}$ \\
Current spin period & $P$ & hr \\
Volume & $V$ & km$^3$ \\ 
Bulk density of material & $\rho$ & kg m$^{-3}$ \\
Friction angle & $\phi$ & deg \\
Mean radius & $R$ & m \\
Cohesive strength & $\hat Y$ & Pa \\
Poisson's ratio & $\nu$ & [-] \\ 
Normalized cohesive strength & $Y$ & [-] \\
Normalized minimum cohesive strength that holds the original shape & $Y^\ast$ & [-] \\
Normalized current spin rate & $\omega_c$ & [-] \\
Slope parameter & $\eta$ & [-] \\
Spin rate parameter & $\omega_0$ & [-] \\
Triaxial ellipsoid's aspect ratio of the semi-intermediate axis to the semi-major axis & $\beta$ & [-] \\
Triaxial ellipsoid's aspect ratio of the semi-minor axis to the semi-major axis & $\gamma$ & [-] \\ 
\bottomrule
\end{tabularx}
\end{center}
\end{table}

\subsection{FEM technique}
\label{Sec:stressAnalysis}
In this section, we explain our FEM technique for analyzing the failure conditions and modes of an irregularly shaped asteroid due to fast rotation \citep{Hirabayashi2014DA, Hirabayashi2015Sphere, Hirabayashi2016Itokawa, Hirabayashi2016Nature}. In this work, we use an FEM solver of ANSYS Mechanical APDL (18.1) licensed by Auburn University's College of Engineering. For the development of an FEM mesh, see Section \ref{Sec:FEMmesh}. We assume that because rotational change due to the YORP effect is on the order of a few million years \citep{Rubincam2000}, the evolution of the internal stress is nearly static \citep{Holsapple2010}. This assumption allows for eliminating the dynamical terms in the stress equations \citep{Holsapple2010}. 

The deformation process consists of elastic and plastic deformation. The elastic model uses Hooke's law, assuming that the material uniformly deforms in any directions. On the other hand, the plastic model in this work uses the associated flow rule to characterize plastic behavior of materials. The associated flow rule defines a relationship between stress and plastic strain based on the yield condition, which is given as
\begin{eqnarray}
d {\bfm \epsilon}_{ij}^p = d \lambda \frac{\partial f}{\partial {\bfm \sigma}_{ij}},
\end{eqnarray}
where $d {\bfm \epsilon}_{ij}^p$ and ${\bfm \sigma}_{ij}$ are the plastic strain change and the stress. $i$ and $j$ are indices that describe the components. $d \lambda$ is a constant value. $f$ is the yield condition and, in the present work, corresponds to the Drucker-Prager yield criterion, which is given in Equation (\ref{Eq:DPcriterion}). Similar to the elastic model, this plastic model describes uniform deformation. In addition, our model assumes no hardening and softening effects, guaranteeing that plastic deformation occurs under constant stress \citep{Chen1988}. 

We consider that a sample asteroid is uniformly rotating at a given spin period. To apply this condition to our FEM simulation, we employed the following method. First, we define a loading acting at each node. Because this loading consists of the gravitational effect and the centrifugal effect, it is necessary to consider the volume of each tetrahedral mesh element. For simplicity, we split its volume into each edge node equally so that it has a mass concentration. Because of this setup operation, however, the location of the center of mass and the orientation of the principal axes change. Because this offset causes translational forces and rotational torques, the simulation quality becomes significantly low, causing unrealistic stress concentration \citep{Hirabayashi2016Itokawa}. To fix this issue, we remove the residual forces induced by this misalignment. This process allows for avoiding the rotational and translational motion of a sample asteroid in our FEM simulation. 

For a description of the deformation process in space, it is ideal that we do not have any constraints on degrees of freedom. However, such an FEM setting does not allow for solving this problem numerically because there are not enough equations. Thus, the number of constraints on degrees of freedom is a key element to making our FEM simulation realistic as much as possible. Here, we only constrain the translational and rotational motion. To do so, we constrain six degrees of freedom. The first three degrees of freedom are added at the note located at the center of mass to remove the translational motion. The next two degree of freedom are given at one of two nodes located along the minimum principal axis to remove the rotation motion along the maximum and intermediate principal axes. The last degree of freedom is provided at one of two nodes located along the minimum principal axis. Note that our FEM technique uses the body-fixed frame. 

To describe the regions that experience plastic deformation, we use the stress ratio, a ratio of the current stress to the yield stress, which is defined by the Drucker-Prager criterion. When the stress ratio is unity, materials at this location should experience plastic deformation. On the other hand, when the stress ratio does not have a unity value, plastic deformation does not occur.   

Using these techniques, we conduct FEM simulation to determine the failure conditions and modes of the asteroids. We use the following iteration process. First, we choose a relatively high cohesive strength as an input parameter and conduct FEM simulation, while the friction angle is fixed at 35 deg through all the iteration processes. Then, we check if there are plastically deformed regions in the considered asteroid. If we do not observe them, we decrease the input value of the cohesive strength and conduct another FEM simulation. We iterate this process until we observe plastic deformation in the sample asteroid.  

We note that the associated flow model used in this analysis may be ideal; however, because we consider the structural failure process of an asteroid due to quasi-static spin-up, our plastic can reasonably predict the locations of the failed regions \citep{Hirabayashi2015internal, Hirabayashi2015Sphere, Hirabayashi2016Nature}. We also mention that while our study assumes the uniformly distributed material condition, it may be possible that the deformation mode may be controlled by the internal structure. In fact, the internal structure distribution have been discussed by many earlier works (e.g., \cite{Hirabayashi2015internal}, \cite{Sanchez2016}, \cite{Zhang2017}, and \cite{Bagatin2018}). Such a detailed analysis will be a subject of our future work. 

\subsection{FEM mesh of an irregularly shaped body}
\label{Sec:FEMmesh}
The quality of an FEM mesh controls the numerical convergence\footnote{Our FEM simulation uses a criterion in which the $L2$ norm of a difference between an actual load and a restored load based on deformation becomes within $0.01 \%$ of the actual load. Note that the actual load is the set of forces acting on the defined nodes that we input based on the effects of gravity and centrifugal forces; on the other hand, the restored load is computed from the solution of deformation \citep{ANSYSThr}. We observe that if the numerical convergence occurs correctly, the convergence rate during each iteration with a constant load step is almost constant, and the $L2$-norm value constantly becomes close to its threshold. If we do not observe this behavior, we do not accept the solution as a proper result.} when an asteroid experiences plastic deformation in calculation. In this section, we describe how to create FEM meshes of the considered asteroids. We use a 10-node tetrahedron FEM mesh. Here, we introduce two mesh-generation techniques. The first technique generates a low-resolution FEM mesh (Technique A), while the second technique produces a high-resolution FEM mesh (Technique B). 

We introduce Technique A, which uses TetGen, a publicly available mesh generator \citep{Si2015}, and ANSYS. The FEM meshes of most of the sample asteroids are generated using this technique. First, we start by generating a 4-node FEM mesh from a polyhedron shape model. Second, we convert this mesh to a 10-node FEM mesh, finding the middle points between the nodes on the edges of each element. A key issue is that TetGen and ANSYS use different measurements to check the quality of an FEM-mesh tetrahedron. While TetGen uses a radius-edge ratio, which is the ratio of the smallest length of the edge to the radius of the circumferential sphere of the tetrahedron, ANSYS applies the edge angle, an angle between two edges of the tetrahedron. This difference induces some inconsistencies; therefore, we refine the quality of the derived FEM mesh using ANSYS. In this work, we define the tolerance value of the edge angle on ANSYS as 165 deg. Finally, we finally add a node at the center of mass of the considered asteroid.

As seen in the supplemental text (Text S1), however, we have difficulties in obtaining the stress ratio distributions of 1996 HW1 and Nereus because of the quality of their FEM meshes obtained by Technique A. A strength of TetGen is that there is a function that automatically changes the mesh resolution in volume, resulting in a high-quality FEM mesh with a fewer number of mesh elements than the same-quality mesh developed at uniform resolution \citep{Si2015}. However, this function results in a difficulty in generating mesh elements in the neck region of 1996 HW1 and a low mesh quality in the central region of Nereus. 

We apply Technique B to avoid these issues to generate the FEM meshes of these asteroids. In this technique, we only use ANSYS. We input the polyhedron shape models of these asteroids to ANSYS and directly generate a 10-node FEM mesh. Then, we add one node at the center of mass manually. We confirm that simulation convergence improves for these asteroids. We also find that even when we use two meshes developed by these techniques, the solutions are comparable under the same physical condition (Text S1). We note, however, that the size of a mesh generated by Technique B is higher than that by Technique A, significantly increasing computational burden.  

\section{Failure mode diagram}
\label{Sec:FMD}
This section explains how we evaluate the failure modes and conditions of the asteroids. If an asteroid experiences structural failure at a low spin rate, $Y^\ast$ of this asteroid should be low because the centrifugal force is small. On the other hand, if this asteroid structurally fails at a high spin rate, $Y^\ast$ should be high to keep its original shape. This fact implies that the failure mode depends on the spin rate, and $Y^\ast$ is a function of $\omega$. To describe the variation in the failure mode, we use a failure mode diagram (FMD), a technique for identifying a failure mode at a given spin rate \citep{Hirabayashi2016Itokawa}. The present study computes the FMDs for all the 24 asteroids; the main text shows the FMDs of four selected asteroids (Itokawa, Geographos, 2008 EV5, and Golevka), and the FMDs of the other asteroids are given in the supplemental information. 

We discuss how to read the FMD. Here, as an example, we use Figure \ref{Fig:Itokawa}\textbf{c}, which describes the FMD of Itokawa. The black line with markers indicates $Y^\ast$ derived from our FEM analysis, while the blue dashed line is a function fitting with our results. If the actual cohesive strength is within the gray region, Itokawa can physically exist. On the other hand, if the cohesive strength is outside this region, Itokawa should structurally fail. Therefore, the white region is considered to be a region in which the cohesive strength of Itokawa violates the spin and shape conditions.  

We divide the trend of $Y^\ast$ into two regimes by the black dot-dashed line (again, see Figure \ref{Fig:Itokawa}\textbf{c}). The left-hand side from the black dot-dashed line is a compression-dominant region, which does not induce structural failure significantly because $Y^\ast$ is nearly zero. On the other hand, to the right of the black dot-dashed line is a tension-dominant region in which the failure model becomes significant. Because $Y^\ast$ monotonically increases as shown in Figure \ref{Fig:Itokawa}\textbf{c}, Itokawa has to have a certain amount of cohesive strength to keep its original shape at a given spin rate. 

To describe $Y^\ast$, we introduce two parameters, $\eta$ and $\omega_0$. In the compression-dominant region, $Y^\ast$ is negligible because tension-driven failure is not a critical factor. In the tension-dominant region, we assume $Y^\ast$ to be a function of a normalized spin rate, $\omega$, which is given as
\begin{eqnarray}
Y^\ast = \eta (\omega^2 - \omega_0^2). \label{Eq:Yast}
\end{eqnarray}
In this equation, $\omega_0$ is the normalized spin rate at which the peak centrifugal stress first balances the local gravity, which is dependent on the shape. On the other hand, $\eta$ is the parameter that describes how quickly the peak centrifugal stress increases with spin rate. If the cross-section perpendicular to the minimum principal axis is small, $\eta$ becomes larger. For Itokawa, because of its contact binary feature, we expect that this asteroid should have a higher value than other asteroids. We determine $\omega_0$ and $\eta$ by fitting Equation (\ref{Eq:Yast}) to our FEM calculations.  

\section{Results}
This section discusses how asteroids structurally fail at different spin rates. As discussed in Section \ref{Sec:FMD}, we will show the failure modes of four selected asteroids representing different shapes. For the C shape, we introduce the FMD of Itokawa. For the E shape, we describe the FMD of Geographos. For the S shape, we indicate the FMD of 2008 EV5. For the N shape, we give the FMD of Golevka. The FMDs of other shapes are described by $\omega_0$ and $\eta$ in Table \ref{Table:AsteriodAnalysis} and are displayed in the supplemental information. 

In the following, to describe the structurally failed regions clearly, we will compute the bimodal stress ratio distribution, i.e., failed regions described in yellow (a unity stress ratio) and non-failed regions given in green (a non-unity stress ratio). To create the bimodal stress ratio distribution, our visualization method defines each triangular tile, or a face, in the same color. Thus, this method uses the node solutions from our FEM simulation and sorts them out. If there is at least one node on a triangular tile, this method decides that this tile is considered to be failed structurally. We note that this process causes the plastic failure region to look wider, depending on the size of a tetrahedron element. However, because our focus is on determining the failure condition and mode, this visualization does not affect our results. 

\subsection{Contact binary objects}
\begin{figure}[ht]
  \centering
  \includegraphics[width=\textwidth]{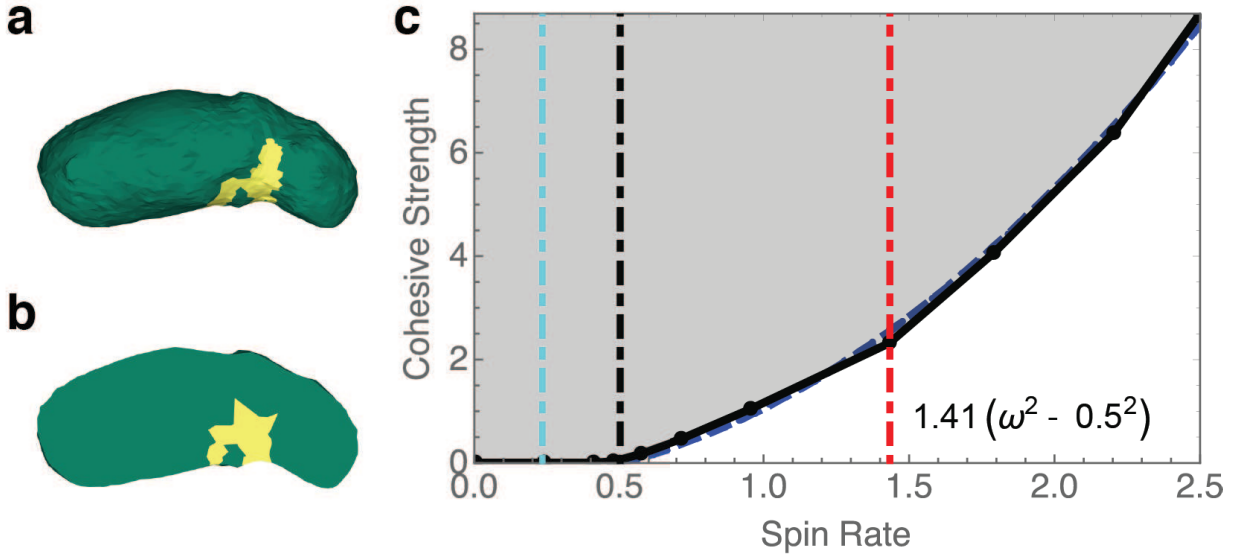}
  \caption{Failure mode and FMD of Itokawa. The failure mode is obtained by solving a case with a cohesive strength equal to $Y^\ast$ at a given spin rate. Figures \ref{Fig:Itokawa}\textbf{a} and \ref{Fig:Itokawa}\textbf{b} give the bimodal stress ratio distribution on the surface and across the cross-section. The yellow regions are failed regions, while the green regions are non-failed regions. The spin axis corresponds to the vertical direction. Figures \ref{Fig:Itokawa}\textbf{c} indicates the FMD. The black solid line with markers shows the distribution of $Y^\ast$ computed by the plastic FEM analysis. The blue dashed line is a fitting function described by Equation (\ref{Eq:Yast}), and the displayed equation is this fitting function. The gray area represents the region in which Itokawa can keep its current shape. The black dot-dashed line describes $\omega_0 = 0.5$. The cyan dot-dashed line gives the normalized current spin rate, $\omega_c$, while the red dot-dashed line indicates that in Figures \ref{Fig:Itokawa}\textbf{a} and \ref{Fig:Itokawa}\textbf{b}.}
  \label{Fig:Itokawa} 
\end{figure}

The failure mode of the C-shape asteroids is characterized as the internal failure of their necks. Figure \ref{Fig:Itokawa} describes the failure mode of Itokawa \citep{Hirabayashi2016Itokawa}. Figures \ref{Fig:Itokawa}\textbf{a} and \ref{Fig:Itokawa}\textbf{b} show the distribution of the stress ratio on the surface and across the cross section at a normalized spin rate of 1.43. The cohesive strength is $Y^\ast = 2.3$\footnote{The dimensional cohesive strength can be computed multiplying the values in the sixth column in Table \ref{Table:AsteriodAnalysis}.}. Figure \ref{Fig:Itokawa}\textbf{c} shows the FMD of Itokawa. We show the spin rate condition used for the simulation above (the red dot-dashed line) and the normalized current spin rate, which is given as $\omega_c$ in Table \ref{Table:AsteriodAnalysis} (the cyan dot-dashed line). The bottom-right equation indicates Equation (\ref{Eq:Yast}) for Itokawa, providing $\eta$ and $\omega_0$ as 1.41 and 0.50, respectively (Table \ref{Table:AsteriodAnalysis}). $\omega_0$ of this body is 0.5, which is given by the black dot-dashed line. As discussed above, at a spin rate less than $\omega_0$, $Y^\ast$ is nearly zero. On the other hand, if the spin rate is higher than $\omega_0$, $Y^\ast$ is described in Equation (\ref{Eq:Yast}). 

We also calculate the failure modes of other C-shape asteroids; see $\omega_0$ and $\eta$ in Table \ref{Table:AsteriodAnalysis} and the failure modes in Figures S.4 through S.8 in the supplemental information. Again, this shape includes Eros, Toutatis, Mithra, Castalia, and 1996 HW1, as well as Itokawa. $\eta$ of this shape ranges from 0.90 to 8.10, giving a mean value of 2.42. 1996 HW1 has a remarkably high $\eta$ value of 8.10. On the other hand, $\omega_0$ ranges between 0.50 and 0.73, and thus the mean value is 0.60. A common feature of the failure mode in this shape class is that the YORP-driven spin-up eventually induces structural failure at their narrow cross sections. When an asteroid in this shape spins up along the maximum principal axis, the centrifugal force increases along the minimum principal axis. Because the cross section becomes smaller on the narrow neck, tensile stress acting on that section grows rapidly. Thus, the C-shape asteroids become sensitive to structural failure at a low spin rate. This mode leads to a breakup into two components. 

\subsection{Elongated bodies}
\begin{figure}[ht]
  \centering
  \includegraphics[width=\textwidth]{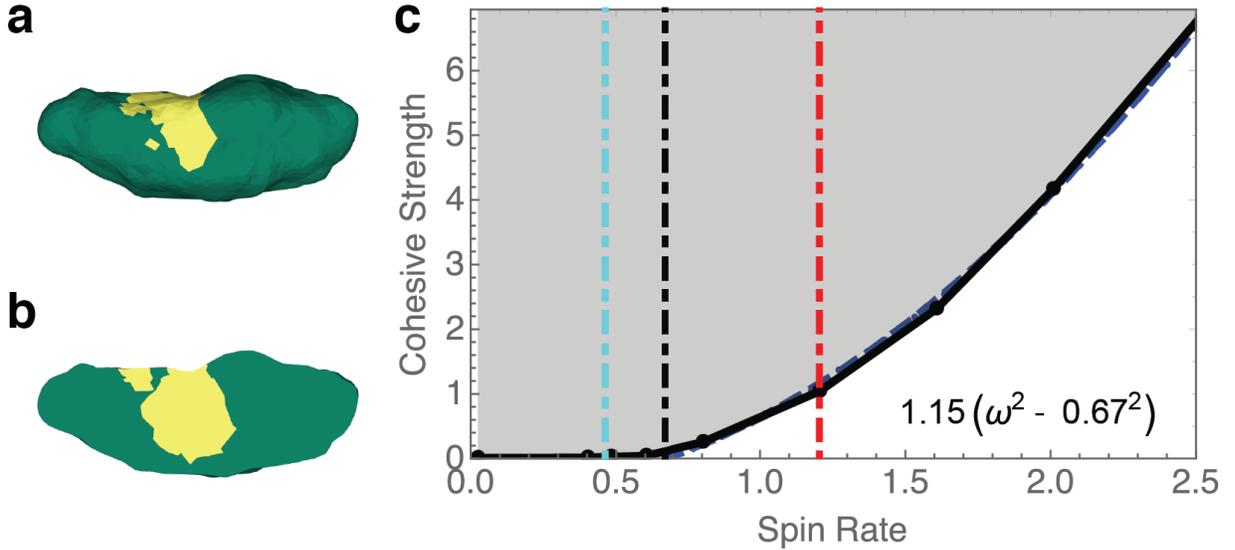}
  \caption{Failure mode and FMD of Geographos. The format of this plot is similar to Figure \ref{Fig:Itokawa}. Figures \textbf{a} and \textbf{b} indicate the stress ratio distribution when $Y^\ast = 1.04$ at $\omega = 1.21$. Figure \textbf{c} shows the FMD of this asteroid. The black dot-dashed line gives $\omega_0 = 0.67$. The cyan dot-dashed line describes $\omega_c = 0.45$, while the red dot-dashed line indicates $\omega = 1.21$.}
  \label{Fig:Geographos} 
\end{figure}

The E shape includes Geographos, Bacchus, Nereus, and 1992 SK. Figure \ref{Fig:Geographos} shows the stress ratio distribution and the FMD of Geographos. Earlier works have shown that the shape of this object might result from a tidal effect during planetary flyby \citep{Bottke1999, Walsh2014}. Geographos is highly elongated but does not have a clear neck structure. The failure mode of this asteroid may be similar to that of the C shape asteroids. However, because there is no distinctive narrow neck, the failed region is located at the middle of the body (Figures \ref{Fig:Geographos}\textbf{a} and \ref{Fig:Geographos}\textbf{b}). This failure mode is attributed to the stress condition on a plane perpendicular to the minimum principle axis around the center of mass of the body. 

We also discuss the FMD of Geographos. The fitting function of $Y^\ast$ at a spin rate higher than $\omega_0 = 0.67$ is given as $1.15 (-0.67^2 + \omega^2)$, leading to $\eta = 1.15$ and $\omega_0 = 0.67$ (Table \ref{Table:AsteroidList}). It is found that $\omega_0$ of this object is higher than that of Itokawa, while $\eta$ of this object is smaller than that of Itokawa. This contrast results from whether or not there is a neck. Because Geographos has no clear neck feature, it does not experience stress concentration that Itokawa has in the neck region. Thus, this asteroid can hold its structure at a higher spin rate than Itokawa, leading to a higher value of $\omega_0$. Also, even when Geographos experiences tension, it does not need a high cohesive strength at a high spin rate, which results in a lower value of $\eta$. 

We show other E-shape asteroids (Table \ref{Table:AsteriodAnalysis} and Figures S.9 through S.11 in the supplemental information). These objects have $\eta$ ranging between 0.33 and 1.15 (Table \ref{Table:AsteriodAnalysis}). The mean value of $\eta$ of the E-shape asteroids is 0.66, which is lower than that of the C shape (2.42). $\omega_0$ is between 0.67 and 0.87, and its mean value is 0.74, which is higher than that of the C shape (0.60). The failure modes of these asteroids are similar to the failure mode of Geographos; commonly, they experience structural failure around the middle of their structure, which induces a breakup at a higher spin rate than the C shape asteroids do.  

We finally point out that 1992 SK may have a different failure mode (Figure S.11). At a spin rate higher than $\omega_0 = 0.87$, the failure region spreads from the middle of the body to the thicker end, implying that the shape would structurally stretch in the left direction. Therefore, instead of breaking up, this asteroid would eventually have material shedding in that way. We suspect that 1992 SK is an example of an asteroid that could have a non-breakup failure mode due to its topological deviation from an elongated shape.

\subsection{Spheroidal objects}
In this section, we discuss the failure modes of the S-shape asteroids. An earlier work showed structural failure at the center of the body at a high spin rate \cite{Hirabayashi2015Sphere}. This deformation mode consists of vertically inward deformation at high latitudes and horizontally outward deformation at low latitudes \citep{Hirabayashi2014DA, Scheeres2016Bennu}. In the case when materials were uniformly distributed, this failure mode was also observed by two independent Soft-Sphere Discrete Element Models \citep{Hirabayashi2015internal, Zhang2017}. 

\begin{figure}[ht]
  \centering
  \includegraphics[width=\textwidth]{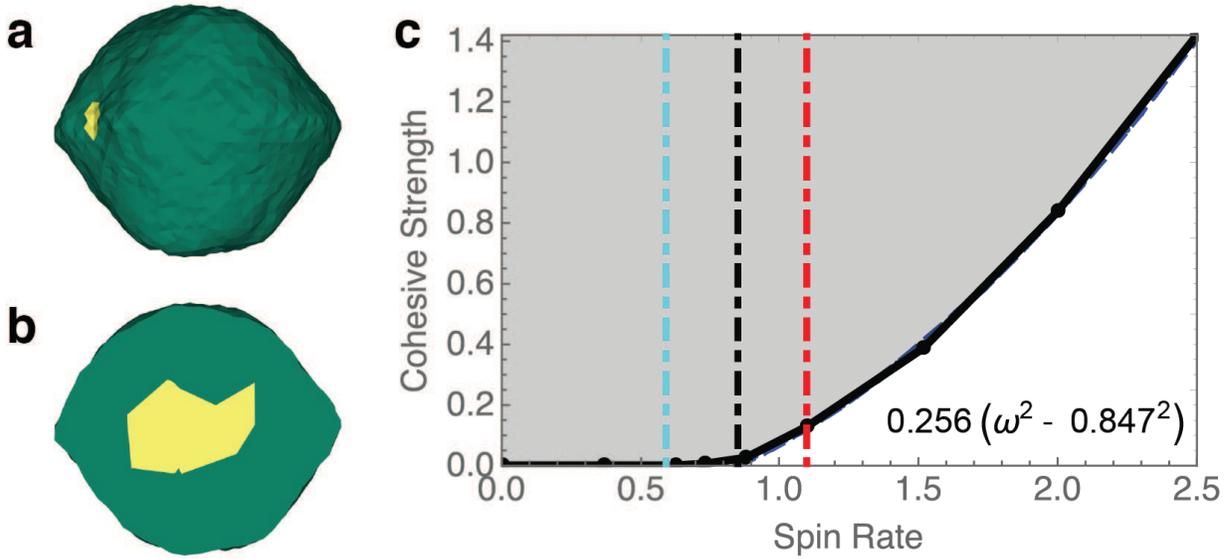}
  \caption{Failure mode and FMD of 2008 EV5. The format of this plot is similar to Figure \ref{Fig:Itokawa}. Figures \ref{Fig:EV5}:\textbf{a} and \textbf{b} display the stress ratio distribution in the case of $Y^\ast = 0.13$ at $\omega = 1.10$. Figure \ref{Fig:EV5}\textbf{c} provides the FMD. The black dot-dashed line shows $\omega_0 = 0.85$. The cyan dot-dashed line describes $\omega_c = 0.59$, while the red dot-dashed line shows the case of Figure \ref{Fig:EV5}\textbf{a} and \textbf{b}.}
  \label{Fig:EV5}  
\end{figure}

Figure \ref{Fig:EV5} shows the failure mode and FDM of 2008 EV5. We note that this object has a concave feature in the equatorial region, which might result from an impact cratering event \citep{Busch2011} or detachment of a small chunk \citep{Tardivel2018}. However, because this feature is small compared to the entire volume, we consider this object to be in the S shape. Figures \ref{Fig:EV5}\textbf{a} and \textbf{b} describe the stress ratio distribution when $Y^\ast = 0.13$ at $\omega = 1.10$. While we observe failed regions around the concave feature, the majority of the subsurface region is intact, and the central region experiences structural failure. This feature is consistent with the results from earlier theoretical and numerical works \citep{Hirabayashi2014DA, Hirabayashi2015Sphere, Hirabayashi2015internal, Scheeres2016Bennu, Zhang2017}. 

Figure \ref{Fig:EV5} indicates the FMD of 2008 EV5. The fitting function of $Y^\ast$ is described as $0.26 (-0.85^2 + \omega^2)$; therefore, $\omega_0$ and $\eta$ are obtained as 0.85 and 0.26, respectively. We find that $\omega_0$ of this object is higher than that of the C and E-shape asteroids, while $\eta$ of this object is smaller than that of these two shapes. Because the shape is spheroidal, the centrifugal force along the minimum principal axis is limited, compared to other two shape classes. This fact allows $\omega_0$ to become higher. Also, the cross section is wide, so even if the spin rate is high, an asteroid in this type does not need a high cohesive strength to keep the original shape, leading to low $Y^\ast$. 

We also show the failure modes of other S-shape asteroids (Table \ref{Table:AsteriodAnalysis} and Figures S.12. through S.23 in the supplemental information). The derived failure mode of these asteroids is consistent with the mode of 2008 EV5; they have low $\eta$ and high $\omega_0$, compared to other shape classes. 

\subsection{Non-classified objects}
\begin{figure}[ht]
  \centering
  \includegraphics[width=\textwidth]{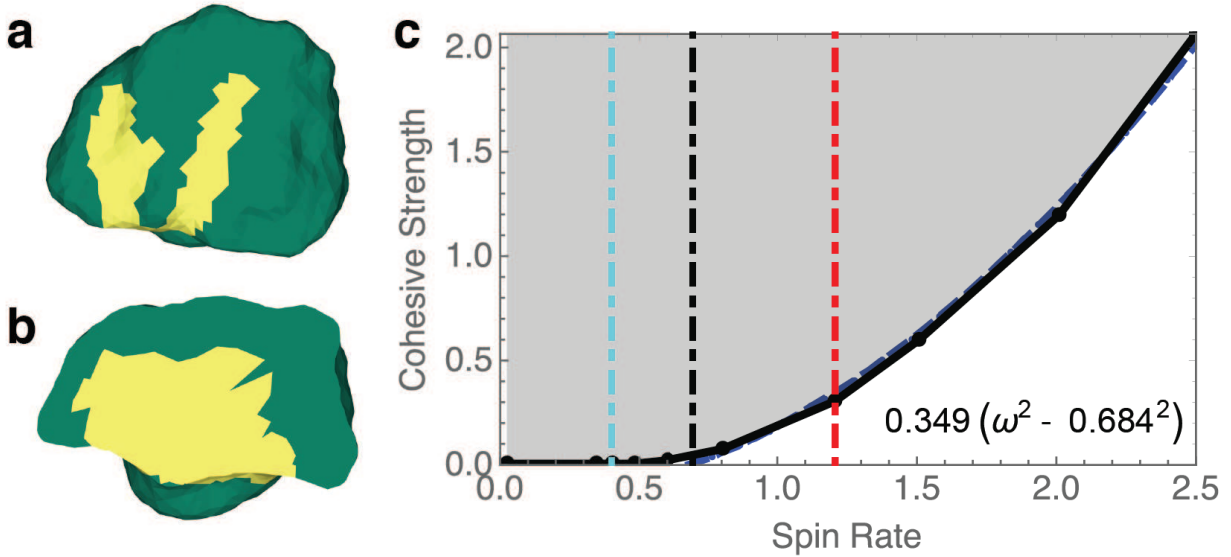}
  \caption{Failure mode and FMD of Golevka.  Figure \ref{Fig:Golevka}\textbf{a} and \textbf{b} indicate the stress ratio distribution in the case of $Y^\ast = 0.31$ at $\omega = 1.21$. \textbf{c} gives the FMD. The black dot-dashed line shows $\omega_0 = 0.68$. The cyan dot-dashed line is the normalized current spin rate, $\omega_c = 0.40$, while the red dot-dashed line is the case of \textbf{a} and \textbf{b}.}
  \label{Fig:Golevka}  
\end{figure}

This section discusses the failure modes of the N-shape asteroid. In our study, only Golevka is categorized into this shape class. This body looks like a tooth and is currently rotating with a spin period of 6.029 hr. Given an assumed bulk density of 2.5 g/cm$^3$, its normalized current spin rate, $\omega_c$, is 0.40 (Table \ref{Table:AsteriodAnalysis}). Figure \ref{Fig:Golevka} provides the stress ratio distribution and the FMD of this object. Figures \ref{Fig:Golevka}\textbf{a} and \ref{Fig:Golevka}\textbf{b} show that the interior fails at a normalized spin rate of 1.21 and $Y^\ast = 0.31$. This failure mode is similar to other shape classes, which had structural failure of the internal structure at a high spin rate. However, because of its tooth-like shape, structural failure appears at the edges of this body, implying that the shape plays an important role in the failure mode. Figure \ref{Fig:Golevka}\textbf{c} displays the FMD of Golevka. $Y^\ast$ is given as $0.35 (-0.68^2 + \omega^2)$; therefore, $\omega_0$ and $\eta$ are 0.68 and 0.35, respectively.

\begin{table}[b]
\begin{center}
\caption{Computationally derived $\eta$, $\omega_0$, and $\omega_c$ for all the asteroids. The fifth column, $P(\omega_0)$, describes the dimensional spin period [hr] at $\omega_0$. The sixth column indicates the dimensional cohesive strength [Pa] at $Y^\ast = 1$. These quantities are obtained by taking into account the size and bulk density of each asteroid given in Table \ref{Table:AsteroidList}.}
\rowcolors{2}{gray!25}{white}
\begin{tabular}{l  l  l  l  l  l  l} 
 \rowcolor{gray!50}
 \toprule
Asteroid System & $\eta$ & $\omega_0$ & $\omega_c$ & $P(\omega_0)$ [hr] & $Y^\ast = 1$ [Pa] \\ 
\hline 
\hline
     {\it Contact binary objects} & & & & & \\
     (433) Eros                  & 1.57 & 0.58 & 0.44 & 4.03 & $1.06 \times 10^2$ \\
     (4179) Toutatis          & 1.11 & 0.55 & 0.026 & 4.39 & $1.96 \times 10^3$ \\
     (4486) Mithra            & 1.40 & 0.73 & 0.036 & 3.28 & $9.37 \times 10^2$ \\
     (4769) Castalia         & 0.90 & 0.73 & 0.59 & 3.33 & $3.85 \times 10^2$ \\
     (8567) 1996 HW1              & 8.10 & 0.52 & 0.28 & 5.16 & $8.58 \times 10^2$ \\
     (25143) Itokawa                 & 1.41 & 0.50 & 0.24 & 5.73 & $1.73 \times 10^1$ \\ 
     & & & & & \\
     {\it Elongated objects} & & & & & \\
     (1620) Geographos   & 1.15 & 0.67 & 0.45 & 3.60 & $2.16 \times 10^3$ \\ 
     (2063) Bacchus       & 0.73 & 0.67 & 0.16 & 3.57 & $1.30 \times 10^2$ \\ 
     (4660) Nereus           & 0.43 & 0.75 & 0.16 & 3.21 & $3.64 \times 10^1$ \\
     (10115) 1992 SK               & 0.33 & 0.87 & 0.33 & 2.97 & $3.31 \times 10^2$ \\ 
     & & & & & \\
     {\it Spheroidal objects} & & & & & \\
     (1580) Betulia                   & 0.37 & 0.83 & 0.39 & 2.89 & $9.52 \times 10^3$ \\
     (2100) Ra-Shalom            & 0.33  & 0.91 & 0.12 & 2.65 & $1.70 \times 10^3$ \\ 
     (29075) 1950 DA                 & 0.35 & 0.78 & 1.38 & 3.08 & $5.52 \times 10^2$ \\
     (33342) 1998 WT24            & 0.32 & 0.95 & 0.65 & 2.54 & $5.65 \times 10^1$ \\
     (52760) 1998 ML14            & 0.27 & 0.76 & 0.16 & 3.16 & $3.22 \times 10^2$ \\
     (66391) 1999 KW4    & 0.30 & 0.99 & 0.98 & 2.74 & $3.50 \times 10^2$ \\
     (101955) Bennu                  & 0.28 & 0.92 & 0.79 & 3.70 & $2.01 \times 10^1$ \\
     (136617) 1994 CC    & 0.25 & 0.90 & 0.99 & 2.94 & $8.73 \times 10^1$ \\
     (153591) 2001 SN263       & 0.24 & 0.87 & 1.06 & 4.20 & $4.21 \times 10^2$ \\   
     (162421) 2000 ET70          & 0.25 & 0.72 & 0.27 & 3.73 & $1.07 \times 10^3$ \\     
     (185851) 2000 DP107       & 0.28 & 0.95 & 1.17 & 3.54 & $7.45 \times 10^1$ \\
     2002 CE26                         & 0.30 & 1.00 & 1.22 & 4.00 & $5.08 \times 10^2$ \\
     2008 EV5                           & 0.26 & 0.85 & 0.59 & 2.60 & $7.75 \times 10^1$ \\ 
     & & & & & \\
     {\it Irregularly shaped objects} & & & & & \\
     (6489) Golevka                  & 0.35 & 0.68 & 0.40 & 3.52 & $9.20 \times 10^1$ \\
\bottomrule
\end{tabular}
\label{Table:AsteriodAnalysis}	
\end{center}
\end{table}

\subsection{Correlation of the shape and failure conditions}
We discuss how the shape of an asteroid influences the failure condition, using $\omega_0$ and $\eta$. As discussed in Section \ref{Sec:FMD}, these parameters are dependent on the shape. We compare these quantities from our results with those from a well-accepted averaging theory for a uniformly rotating triaxial ellipsoid \citep[e.g.][]{Holsapple2001, Holsapple2004, Holsapple2007, Holsapple2010, Sharma2009, Rozitis2014, Hirabayashi2014Biaxial}. If the irregularity of the shape does not control the failure conditions, these techniques should give consistent results. Here, to describe the elongation, we define $\beta$ and $\gamma$. $\beta$ is the ratio of the semi-intermediate axis to the semi-major axis, while $\gamma$ is that of the semi-minor axis to the semi-major axis. 

\begin{figure}[ht]
  \centering
  \includegraphics[width=5in]{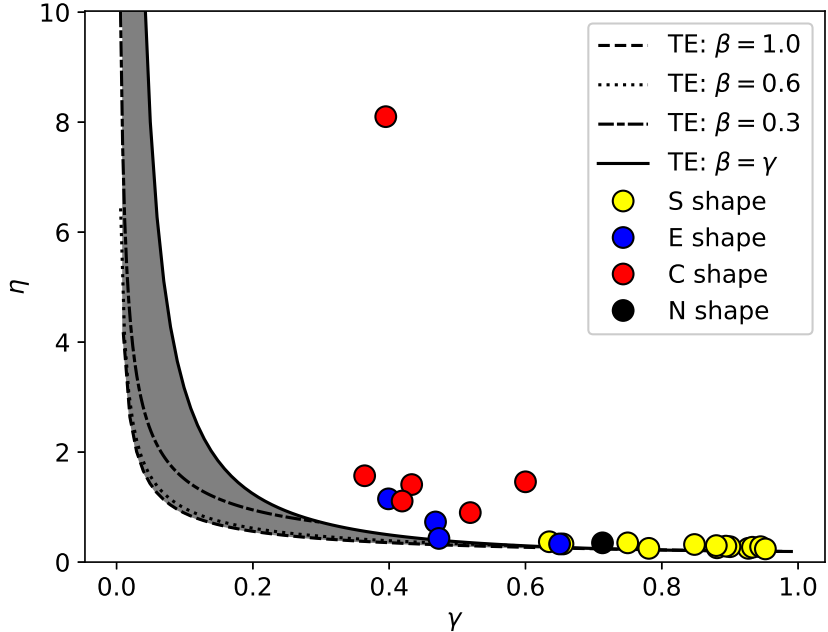}
  \caption{Distribution of $\eta$ as a function of the elongation. The yellow circles describe the S-shape objects, the blue circles indicate the E-shape objects, the red circles show the C-shape objects, and the black circle draws the N-shape object. The blue line indicates the theoretical prediction.}
  \label{Fig:eta}  
\end{figure}

\begin{figure}[ht]
  \centering
  \includegraphics[width=5in]{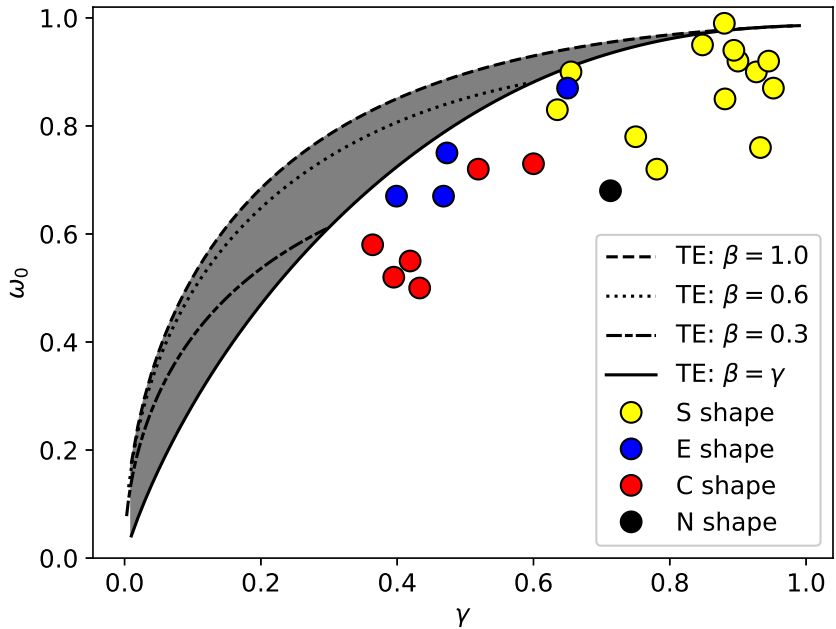}
  \caption{Distribution of $\omega_0$ as a function of the elongation. The format follows Figure \ref{Fig:Omega0}. }
  \label{Fig:Omega0}  
\end{figure}

Figures \ref{Fig:eta} and \ref{Fig:Omega0} describe the values of $\eta$ and $\omega_0$, respectively, as a function of $\gamma$. The circles describe the results from our FEM study (see Table \ref{Table:AsteriodAnalysis}). On the other hand, the lines show the triaxial ellipsoid model with different $\beta$s and $\gamma$s. If $\gamma = 1$, the asteroid is oblate, giving the upper bound of $\omega_0$ and the lower bound of $\eta$. If $\beta = \gamma$, it is a biaxial ellipsoid, providing the lower bound of $\omega_0$ and the upper bound of $\eta$. If this theoretical model can describe the failure conditions and modes, all the FEM results should be inside the regions sandwiched by these boundaries, which are given in gray. 

We briefly explain how to derive $\omega_0$ and $\eta$ of the triaxial ellipsoid model. First, we compute the stress components that are averaged over the entire volume. Because of the axisymmetric shape, this averaging technique makes the shear stress components zero. Second, substituting the stress components into Equation (\ref{Eq:DPcriterion}), we obtain the FMD for this ideal shape. Third, we numerically determine $\omega_0$ and $\eta$. Because of complexity of Equation (\ref{Eq:DPcriterion}), it is difficult to determine these quantities analytically. Thus, we develop a numerical algorithm that approximately finds these quantities. In this algorithm, we choose two normalized spin rates in the tension region and use Equation (\ref{Eq:DPcriterion}) to compute the normalized cohesive strengths at these spin rates. Then, we compute $\omega_0$ and $\eta$ such that these two points are on the parabolic curve given in Equation (\ref{Eq:Yast}).  

The results show that our FEM analysis and the theoretical model give different failure conditions. The variations in $\eta$ indicate dependence of the failure condition on the shape (Figure \ref{Fig:eta}). The S-shape objects are almost on the theoretical prediction curve, while the elongated objects are slightly off the line. The C-shape objects are highly deviated from that curve, indicating that the neck region of each asteroid has to hold strong tension to keep the original shape unchanged. In fact, 1996 HW1 has the highest of $\eta$ among the considered asteroids, i.e., $> 8$ (see Table \ref{Table:AsteriodAnalysis}), and has an extremely narrow neck (see Figure S8). Golevka, the N-shape asteroid, has a trend similar to the spheroidal objects. 

We also observe that the $\omega_0$ distributions of our results are different from those of the triaxial ellipsoid model (Figure \ref{Fig:Omega0}). While all the shape types give deviation from the triaxial ellipsoid model prediction, we emphasize that the S-shape bodies are highly deviated from it. We attribute this trend to the fact that global irregularity in shape changes the loading condition, causing these asteroids to have different internal failure modes and conditions. 

\section{Discussion}
Our results show several key insights into the failure modes of asteroids due to quasi-static spin-up. Importantly, we find that although the failure modes of irregularly shaped bodies are different, each shape type has its own distinctive features. In this section, we discuss interpretations of our results into the shape evolution of asteroids. 

\subsection{Influence of shapes on failure modes }
Our results indicated that the ``global" shape is a strong contributor to the failure modes and conditions. We divided the sample asteroids into four subjective shape categories. Each category had different failure features. For the S-shape objects, structural failure starts from the internal part, spreading over the entire structure. Most of the E-shape objects have a failure mode in which the failed region spreads over the middle of the body while the edge regions are usually not failed. The failure mode of the C-shape objects is similar to that of the E-shape objects; however, the failed regions appear at their narrow necks. Because the area across the neck is smaller than other areas, the stress is usually concentrated on the neck, causing an asteroid in this shape type to be more sensitive to structure failure than other asteroids. The failure mode of the N-shape object was different from that of an asteroid in other categories. 

We also compared $\eta$ and $\omega_0$ from our FEM technique with those from the volume-averaging technique, concluding that the results from these two techniques were not consistent. There are two reasons for this inconsistency. First, it is too ideal to use a uniformly rotating ellipsoid to model the shape of an asteroid. Global topographic variations change the self-gravity forces and the centrifugal forces, changing the stress field. This induces different failure modes and conditions. Second, the triaxial ellipsoid model technique might overestimate $\omega_0$. Again, $\omega_0$ is the value at which the stress becomes tensile and causes the shape to fail without cohesion. This technique takes an average of the stress distribution over the volume, simplifying the problem. However, by doing so, we lose the fact some regions may experience compression while other regions may have tension. This process might produce strong deviation of the triaxial ellipsoid model from our results. We conclude that consideration of the actual shape is critical to determine the failure condition of an irregularly shaped asteroid. 

\subsection{Interpretation into the shape evolution of asteroids}
The failure modes of the C-shape asteroids and the E-shape asteroids give insights into the formation scenario that these asteroids settle into the C shape. For the C-shape asteroids, the narrow neck is the most sensitive to failure, leading to fission. For the E-shape asteroids except for 1992 SK, the middle region first experiences structural failure. Similar to the C-shape asteroids, they are likely to split into two components. Once the body breaks up, the split components could eventually contact one another due to their mutual dynamics. Whether or not this reconfiguration process occurs depends on the mass ratios of the split components \citep{Scheeres2007A, Pravec2010, Jacobson2011, Scheeres2018}. If the mass ratio of the smaller component to the larger component is higher than 0.2, the mutual orbit is unstable, leading to a soft contact of these components to generate a new contact binary configuration. \cite{Hirabayashi2016Nature} confirmed that the bilobate nuclei of 67P satisfied this condition. 

To test this hypothesis, we calculate the predicted mass ratios after fission of the C-shape asteroids and the E-shape asteroids except for 1992 SK (Table \ref{Table:Massratio}). To derive the mass ratios, we first find planes across the simulated failure regions of these asteroids. Then, we numerically cut them through the planes and computed their mass ratios. For the C-shape asteroids, the cutting planes are aligned along the narrow necks. For the E-shape asteroids, on the other hand, the cutting plans are located almost in the middle of their structure.  

We obtain that all the asteroids have mass ratios higher than 0.2, indicating that after fission, the split components of these asteroids have unstable mutual orbits and eventually contact each other with a low collision velocity. Thus, a possible formation process is that after asteroids become elongated due to some processes such as the YORP-driven spin-up \citep{Holsapple2010} or the tidal effect \citep{Bottke1999, Walsh2014}, they eventually split into multiple components and re-accrete due to soft-contact, having contact binary configurations. We note that another possible scenario is that tidal disruption could directly turn an E-shape object into multiple C-shape objects. Finally, objects that have topographic deviation from an elongated shape, like 1992 SK, may have a different deformation process.

\begin{table}[ht]
\begin{center}
\caption{Calculated mass ratios of the C-shape asteroids and the E-shape asteroids except for 1992 SK after fission.}
\rowcolors{2}{gray!25}{white}
\begin{tabular}{l  l  l} 
 \rowcolor{gray!50}
 \toprule
Asteroid System & Mass ratio  \\ 
\hline 
     {\it Contact binary bodies} &  \\
     (433) Eros                          & 0.95 \\
     (4179) Toutatis                   & 0.21 \\
     (4486) Mithra                     & 0.70 \\
     (4769) Castalia                  & 0.67 \\
     (8567) 1996 HW1              & 0.51 \\
     (25143) Itokawa                 & 0.24 \\ \\
     {\it Elongated bodies except for 1992 SK} & \\
     (1620) Geographos           & 0.66 \\ 
     (2063) Bacchus                 & 0.63 \\ 
     (4660) Nereus                   & 0.77 \\
\bottomrule
\end{tabular}
\label{Table:Massratio}	
\end{center}
\end{table}

\subsection{Effects of heterogeneous structure on failure modes}
An understanding of the failure modes of irregularly shaped asteroids will provide strong constraints on the internal structure. Our analysis showed that an S-shape asteroid could structurally fail from its central point when it rotates rapidly. This failure mode consists of vertical deformation along the spin axis and outward deformation on the equatorial plane. Thus, the shape becomes more oblate than before as the interior pushes the equatorial surface outward \citep{Hirabayashi2014DA, Hirabayashi2015internal, Scheeres2016Bennu}. 

On the contrary, earlier works argued that the oblateness of an S-shape body could result from landslides on its surface \citep{Walsh2008, Minton2008, Harris2009, Walsh2012, Scheeres2015Land}. \cite{Hirabayashi2015internal} discussed that these different mechanisms would both be possible due to various internal structures. If the structure were uniform, the central point would experience the failure mode first. If the body were covered by loosely packed materials, landslides would be a likely contributor to the oblateness. 

On the other hand, \cite{Tardivel2018} proposed that the cohesive strength of the body causes a small, aggregated mass to depart from the equatorial ridge of the body at a rapid rotation state, suggesting that this process could induce the observed concave regions on 2008 EV5 \citep{Busch2011} and 2000 DP107 \citep{Naidu2015}. The ongoing asteroid exploration missions, Hayabusa 2 \citep{Tsuda2013} and  OSIRIS-REx \citep{Lauretta2012}, will be able to shed light on the internal structure based on the surface morphology. We finally note that \cite{Statler2015} hypothesized that the top-shaped asteroids could be less affected by the YORP effect, challenging studies about the spin-up mechanism of S-shape asteroids due to the YORP effect. 

\subsection{Possible numerical artifacts in our FEM technique}
In this section, we introduce possible artifacts in our numerical FEM technique. We first consider a possibility that while the numerical convergence criterion is satisfied in each iteration, the numerically obtained solution may not be equal to the true solution. Investigations of this issue may require a numerous amount of simulation cases and thus a tremendous amount of time, which may not be ideal. To address this issue, however, we refer to \cite{Hirabayashi2015Sphere}, who analyzed the failure condition and mode of a spherical object rotating at a constant spin rate. He compared the theoretical solution and the FEM solution. It was found that while small variations in the stress distribution were observed in local areas, these solutions were consistent. Therefore, while the possibility of this issue in all the cases cannot be ruled out, because our present FEM framework is based on \cite{Hirabayashi2015Sphere}, we consider our technique to provide reasonable solutions. 

Another point to be addressed is whether the resolution of the polyhedron shape model affects the FEM solutions. This issue is partially answered by \cite{Hirabayashi2015Sphere}, in which the theoretical model was consistent with the ANSYS calculation even if the geometry of the FEM mesh was not entirely identical to that of the theoretical sphere because of the artificial resolution. Also, the local topographic features might affect the FEM solution. The higher the resolution of the polyhedron shape model becomes, the more the shape model accounts for small topographic variations such as the detailed shapes of boulders. However, because the failure conditions on a global scale are controlled by the global distributions of the gravity force and the centrifugal force, the small topographic variations do not globally contribute to structure failure much. 

\section{Conclusion}
In this study, we discussed how the shape of an irregularly shaped body would evolve due to the YORP effect, using a finite element model technique. Assuming that materials in objects were homogeneously distributed, we investigated the YORP-driven failure conditions and modes of 24 asteroids ($<$ 40km in diameter) observed at high resolution. Our results showed that the irregular shape of an asteroid is a critical factor that controls the failure mode and condition, pointing out a limited capability of the well-accepted averaging technique. We used a subjective shape classification that divided asteroids into four shape classes: spheroidal objects, elongated objects, contact binary objects, and non-classified objects. We found distinctive trends of the failure mode for each shape type, shedding light on the shape formation processes of asteroids. Structural failure of the spheroidal objects always started from the interior, while the elongated objects had structural failure in the middle of their structure. The contact binary objects, on the other hand, experienced structural failure around their neck region. Also, the failure conditions were highly controlled by these shape features. Further investigations will give constraints on the formation processes of asteroids. 

\section{Acknowledgements}
M.H. acknowledges support from the faculty startup fund from Department of Aerospace Engineering at Auburn University. The software package ANSYS 18.1, licensed by Samuel Ginn College of Engineering at Auburn University, was used for the stress analyses presented in this paper. M. H. thanks Dr. Shantanu Naidu (JPL), Dr. Marina Brozovi{\'c} (JPL), Dr. Jean-Luc Margot (UCLA), Dr. Tracy Becker (SwRI), Dr. Ellen Howell (U. of Arizona), Dr. Patrick Taylor (Arecibo Observatory), and Dr. Michael Busch (SETI) for shape models and useful discussions. 

\newpage


\end{document}